# Simulation work on Fractional Order PI$^\lambda$ Control Strategy for speed control of DC motor based on stability boundary locus method

N.N. Praboo[#1] , P.K. Bhaba[#2]

[1] *Department of Instrumentation Engineering, Annamalai University, Annamalai Nagar, India*
praboonn@gmail.com
[2] *Process Control Lab, Department of Chemical Engineering, Annamalai University, Annamalai Nagar, India*
pkbhaba@gmail.com

*Abstract*— This paper deals with the design of Fractional Order Proportional Integral (FO-PI$^\lambda$) controller for the speed control of DC motor. A mathematical model of DC motor control system is derived and based on this model fractional order PI$^\lambda$ controller is designed using stability boundary locus method to satisfy required gain margin (GM) and phase margin (PM) of the system. Servo and Regulatory tracking simulation runs are carried out for the speed control of DC motor. The performance of the fractional order PI$^\lambda$ (FO-PI$^\lambda$) controller is compared with Integer Order Relay Feedback Proportional Integral (IO-RFPI) controller. Finally the stability of both control system is considered.

*Keywords*— Fractional order control, PID controllers, DC Motor, Speed control system, Optimization, CRONE.

## I. INTRODUCTION

The use of fractional calculus has gained popularity among many research areas during the last decade. Its theoretical and practical interests are well established nowadays, and its applicability to science and engineering can be considered as an emerging new analytical approach. The introduction of fractional order calculus to conventional controller design extends the scope of added performance improvement.

The classical PI and PID controllers remain the most prevalent controllers for many industrial applications over the past decades. In recent years, fractional order dynamic systems and controllers based on fractional order calculus have gained an increasing attention in control community [2]. This is mainly due to the fact that many real physical systems are well characterized by fractional order differential equations involving non integer order derivatives [3].

The concept of fractional order PI$^\lambda$D$^\mu$ controller which has an integrator of real order $\lambda$ and differentiator of real order $\mu$ is proposed by Podlubny [4]. The transfer function of this controller is given by

$$C(s) = K_p + \frac{K_i}{s^\lambda} + K_d s^\mu \quad \ldots\ldots\ldots\ldots\ldots\ldots\ldots\ldots\ldots(1)$$

The PI$^\lambda$D$^\mu$ algorithm is represented by a fractional integro-differential equation of type as follows

$$u(t) = K_p e(t) + K_i D^{-\lambda} e(t) + K_d D^\mu e(t) \quad \ldots\ldots\ldots\ldots\ldots\ldots(2)$$

Where, D is the integro-differential operator [5], e(t) is the controller input and u(t) is the controller output. Clearly, depending on the values of the orders $\lambda$ and $\mu$, the numerous choices for the controller's type can be made. For instance, taking $\lambda$=1 and $\mu$=1 yields the classical PID controller. Moreover, the selection of $\lambda$=1 and $\mu$=0 leads to the PI controller, $\lambda$=0 and $\mu$=1 gives the PD controller, and also $\lambda$=0 and $\mu$=0 results the P controller.

Classification of dynamic systems according to the order of the plant and the controller can be done as: i) integer order system - integer order controller ii) integer order system - fractional order controller iii) fractional order system - integer order controller and iv) fractional order system - fractional order controller.

In this paper design of fractional order PI$^\lambda$ controller for the speed control of DC motor is made based on stability boundary locus method. In order to get fractional order PI$^\lambda$ controller parameters, the mathematical model of DC motor system is derived. Then, the global stability regions for different values of $\lambda$ in ($K_p$, $K_i$)-plane is obtained. Finally, the controller parameters corresponding to the GM and PM requirements are chosen.

The paper is organized as follows: Section 2 gives a brief description of DC motor speed control system. The mathematical modeling of DC motor speed control system is obtained in Section 3. In Section 4, the fractional PI$^\lambda$ controller design based on the stability boundary locus method is given. The comparative simulation results for the control performance are presented in Section 5. Finally, concluding remarks are given in Section 6.

## II. PROCESS DESCRIPTION

### A. Description of the DC motor speed control system

The functional diagram of speed control of DC motor is shown in Fig. 1. The setup consists of DC motor, chopper driver unit, Opto-coupler sensor, V-MAT card and personal computer. A separately excited linear DC motor is considered for this paper work. The speed of the DC motor is controlled by varying the armature voltage through chopper circuit by varying the PWM pulses.

The chopper circuit is used to convert the pulse width modulated signal from the personal computer through the card into the corresponding voltage signal. The V-MAT card acts as a data acquisition card to interface motor circuit with personal computer. Matlab® Simulink environment is used to monitor and control the speed of the motor from the personal computer.

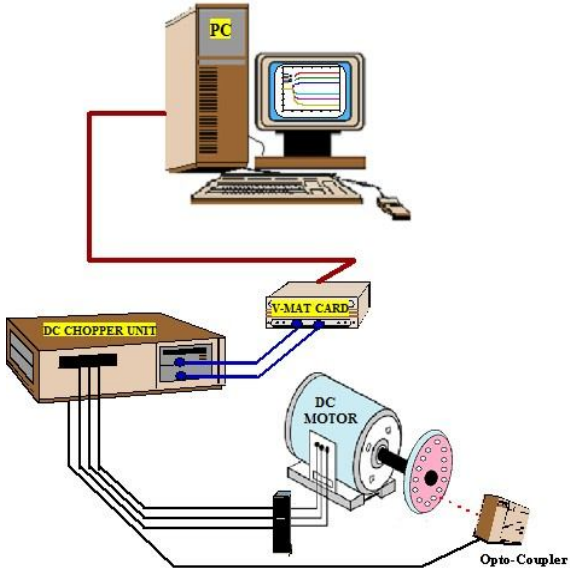

Fig. 1. Experimental setup of a DC motor speed control system.

The DC motor system consists of a corrugated plate in the rotary shaft to transfer speed into pulse, known as Opto-coupler setup. A multifunction VMAT01 interface board including high speed analog to digital converter (ADC) and digital to analog converter (DAC) is interfaced with a PC-AT Pentium 4. The interface card is capable of running the real time control algorithms in Simulink tool of MATLAB platform directly. The obtained voltage signal is processed and the real time control algorithm is carried out by using the VMAT01. The parameters of LLS are shown in Table I.

TABLE I. THE EXPERIMENTAL PARAMETERS OF DC MOTOR SYSTEM.

| | |
|---|---|
| Moment of Inertia of the rotor | J =0.03 kgm$^2$ |
| Maximum Speed of the motor | 1500 rpm |
| Damping (friction) of the mechanical system | b =0.019 Nms |
| $K_b$=$K_T$=K | K =0.1331 |
| Electric Resistance | R = 6Ω |
| Electric Inductance | L = 4.5 mH |

## III. MATHEMATICAL MODEL OF DC MOTOR

### A. Mathematical modelling of DC motor speed control system

In this paper the speed of DC motor is controlled by varying the armature voltage of the motor coil. The pulse width modulated signal from the personal computer is converted into corresponding armature voltage. The armature voltage controls the motor velocity.

The control equivalent circuit of the DC motor by the armature voltage control method is shown in Fig. 2. The mathematical model is derived from the control circuit based on its input, output and inherent parameters of the DC motor.

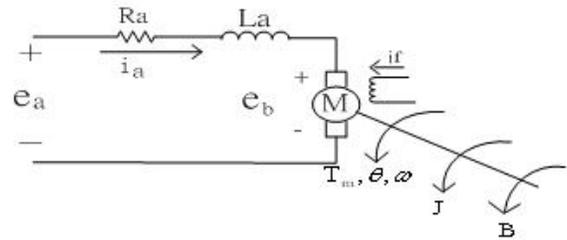

Fig. 2 Control circuit of the DC motor using the armature voltage control

$R_a$ : armature resistance, $L_a$: armature inductance,
$i_a$: armature current, $i_f$: field current
ω: angular velocity of motor,  J: rotating inertial measurement of motor bearing.
$e_a$: input voltage, $e_b$: back electromotive force (EMF),
$T_m$: motor torque, B: damping coefficient

The functional block diagram of DC motor speed control system is shown in Fig. 3.

Because the back EMF $e_b$ is proportional to speed ω directly, then

$$e_b(t) = K_b \frac{d\theta(t)}{dt} = K_b \omega(t) \cdots\cdots\cdots\cdots (3)$$

Making use of the KCL voltage law can get

$$e_a(t) = R_a i_a(t) + L_a \frac{di_a(t)}{dt} + e_b(t) \cdots\cdots (4)$$

From Newton law, the motor torque can obtain

$$T_m(t) = J \frac{d^2\theta(t)}{dt^2} + B \frac{d\theta}{dt} = K_T i_a(t) \cdots\cdots (5)$$

Taking Laplace transform for the above given equations, the equations can be formulated as follows.

$$E_b(s) = K_b \omega(s) \cdots\cdots\cdots\cdots\cdots\cdots\cdots\cdots\cdots\cdots (6)$$

$$E_a(s) = (R_a + L_a s)I_a(s) + E_b(s) \cdots\cdots\cdots\cdots (7)$$

$$T_m(s) = Js\omega(s) + B\omega(s) = K_T i_a(s) \cdots\cdots\cdots (8)$$

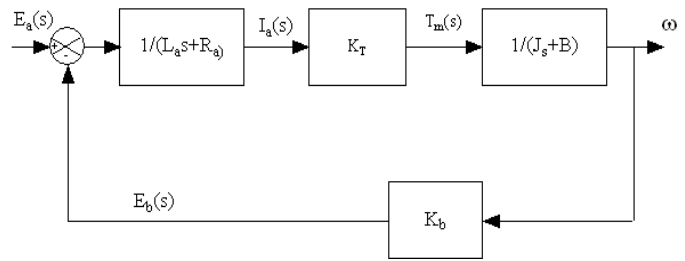

Fig. 3 DC motor armature voltage control system functional block diagram

*B. Transfer model of DC motor speed control system*

The transfer function model is obtained by substituting the experimental DC motor parameters specification given in the table 1.0. The transfer function model is obtained as

$$G(s) = \frac{\omega(s)}{E_a(s)} = \frac{1.01}{0.0021025\,s^2 + 1.367\,s + 1} \quad\cdots\cdots\cdots (9)$$

## IV. FRACTIONAL ORDER CONTROLLER DESIGN

*A. Fractional Order $PI^\lambda$ Control System*

The unity feedback control system consists of a plant $G(s)$ and a controller $C(s)$. In this section, either the plant or the controller is chosen as fractional order. The transfer function of the fractional order plant is expressed in the quasipolynomial format as.

$$G(s) = \frac{N(s)}{D(s)} = \frac{b_n s^{\beta_n} + b_{n-1} s^{\beta_{n-1}} + \ldots\ldots + b_0 s^{\beta_0}}{a_n s^{\alpha_n} + a_{n-1} s^{\alpha_{n-1}} + \ldots\ldots + a_0 s^{\alpha_0}}$$

$$= \sum_{i=0}^{n} b_i s^{\beta_i} \bigg/ \sum_{i=0}^{n} a_i s^{\alpha_i} \quad\cdots\cdots\cdots\cdots (10)$$

where, $\beta_n > \ldots\ldots > \beta_1 > \beta_0 \geq 0$ , $\alpha_n > \ldots\ldots > \alpha_1 > \alpha_0 \geq 0$ , $a_i$ and $b_i$ are arbitrary real numbers.

The transfer function of the fractional order $PI^\lambda$ controller is

$$C(s) = \frac{U(s)}{E(s)} = K_p + \frac{K_i}{s^\lambda} \quad\cdots\cdots\cdots\cdots\cdots (11) .$$

This transfer function is obtained for $K_d = 0$ in (1).

By taking the value of $\lambda$ as 1, the fractional order controller is converted to the classical integer order PI format.

The output of the unity feedback control system is given by

$$y = \frac{C(s)G(s)}{1 + C(s)G(s)} \cdot r \quad\cdots\cdots\cdots\cdots\cdots\cdots (12)$$

Where, $r$ is the reference input and $y$ is the output of the control system. The denominator of (12) represents the *fractional order characteristic quasipolynomial* (FOCQ) of the closed loop system.

*B. Formation of Global Stability Region*

The set of all stabilizing controller parameters is obtained through design from the global stability region in the controller parametric space. Therefore, the designer has the set of all stabilizing controllers for the plant. The literature review in many decades shows many stabilization methods like Stability Boundary Locus method [8, 9], the D-decomposition method [10], the Hermite-Biehler theorem [12], Parameter space approach, etc. In this paper, we make use of the results of [9] which consider the Stability Boundary Locus method. Putting (10) and (11) in (12), the FOCQ is written as

$$P(s) = \sum_{i=0}^{n} \left[ a_i s^{(\alpha_i + \lambda)} + K_p b_i s^{(\beta_i + \lambda)} + K_i b_i s^{\beta_i} \right] = 0$$

$$\cdots\cdots\cdots\cdots (13)$$

Replacing $s = j\omega$ in (13) gives

$$P(j\omega) = \sum_{i=0}^{n} \left[ a_i (j\omega)^{(\alpha_i + \lambda)} + K_p b_i (j\omega)^{(\beta_i + \lambda)} + K_i b_i (j\omega)^{\beta_i} \right] = 0$$

$$\cdots\cdots\cdots\cdots (14)$$

Using the mathematical identity, (14) is written as

$$P(j\omega) = \sum_{i=0}^{n} \left[ \begin{array}{l} a_i(\omega)^{(\alpha_i + \lambda)} \left[ \cos\left\{ (\alpha_i + \lambda)\frac{\pi}{2} \right\} + j\sin\left\{ (\alpha_i + \lambda)\frac{\pi}{2} \right\} \right] \\ + K_p b_i(\omega)^{(\beta_i + \lambda)} \left[ \cos\left\{ (\beta_i + \lambda)\frac{\pi}{2} \right\} + j\sin\left\{ (\beta_i + \lambda)\frac{\pi}{2} \right\} \right] \\ + K_i b_i(\omega)^{\beta_i} \left[ \cos\left\{ (\beta_i)\frac{\pi}{2} \right\} + j\sin\left\{ (\beta_i)\frac{\pi}{2} \right\} \right] \end{array} \right]$$

$$\cdots\cdots\cdots\cdots (15)$$

Equating the real and imaginary parts of $P(j\omega)$ to zero, the real part is obtained as

$$\sum_{i=0}^{n} \left[ \begin{array}{l} a_i(\omega)^{(\alpha_i + \lambda)} \left[ \cos\left\{ (\alpha_i + \lambda)\frac{\pi}{2} \right\} \right] + \\ K_p b_i(\omega)^{(\beta_i + \lambda)} \left[ \cos\left\{ (\beta_i + \lambda)\frac{\pi}{2} \right\} \right] \\ + K_i b_i(\omega)^{\beta_i} \left[ \cos\left\{ (\beta_i)\frac{\pi}{2} \right\} \right] \end{array} \right] = 0 \quad\cdots\cdots (16)$$

and the imaginary part is determined as

$$\sum_{i=0}^{n} \left[ \begin{array}{l} a_i(\omega)^{(\alpha_i + \lambda)} \left[ \sin\left\{ (\alpha_i + \lambda)\frac{\pi}{2} \right\} \right] + \\ K_p b_i(\omega)^{(\beta_i + \lambda)} \left[ \sin\left\{ (\beta_i + \lambda)\frac{\pi}{2} \right\} \right] \\ + K_i b_i(\omega)^{\beta_i} \left[ \sin\left\{ (\beta_i)\frac{\pi}{2} \right\} \right] \end{array} \right] = 0 \quad\cdots\cdots (17)$$

Solving (16) and (17), the controller parameters are obtained by

$$K_p = \frac{1}{Sin\left\{ \lambda \left[ \frac{\pi}{2} \right] \right\}} \frac{\left[ A_1(\omega)B_2(\omega) - A_2(\omega)B_1(\omega) \right]}{\left[ B_1^2(\omega) + B_2^2(\omega) \right]} \quad\cdots\cdots (18)$$

$$K_i = \frac{-\omega^\lambda}{Sin\left\{ \lambda \left[ \frac{\pi}{2} \right] \right\}} \frac{\left[ A_3(\omega)B_2(\omega) - A_4(\omega)B_1(\omega) \right]}{\left[ B_1^2(\omega) + B_2^2(\omega) \right]} \quad\cdots\cdots (19)$$

Using (18) and (19), a stability locus curve is drawn in the $(K_p$-$K_i)$-plane for any value of $\lambda$ ($\omega$ changes from 0 to maximum). Using the test points inside and outside of the curve, the global stability region is obtained.

Considering the DC motor transfer function model in (10), the stability locus curves are obtained for the various values of $\lambda$ in the range of (0, 2). For each curve, the test points are considered and the global stability regions are obtained. The set of global stability regions is shown in the $(K_p$-$K_i)$-plane in Fig. 4.

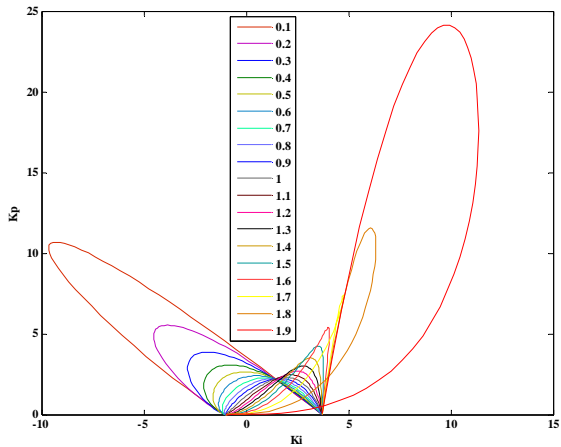

Fig. 4. The global stability regions for different values of $\lambda$. These regions are within the curves.

### C. Selection of Fractional Order $PI^\lambda$ Controller Parameters

From the global stability regions, the designer has flexibility for choosing the controller parameters, $K_p$, $K_i$ and $\lambda$. For the selection, the key idea is to find the controller parameters using the test points in the global stability regions until obtaining the desired open loop GM and PM requirements. A lot of ($K_p$, $K_i$, $\lambda$) values providing these requirements can be determined.

For the DC motor transfer function model in (10), the goal is to obtain the GM of 4.5 dB and PM of 20°. These requirements are provided for only the values of the $\lambda$ in the range of (1.15, 1.25). Therefore, we choose $\lambda$=1.2 and determine the controller parameters as $K_p$=2.5732, $K_i$=1.45204. So the fractional order $PI^\lambda$ controller transfer function is found as

$$G_{CF}(s) = 2.5732 + \frac{1.45204}{s^{1.2}} \qquad \ldots\ldots\ldots\ldots(20)$$

The open loop Bode plot [13] for these values is given in Fig. 5. This figure shows that these values satisfy the constraints mentioned above.

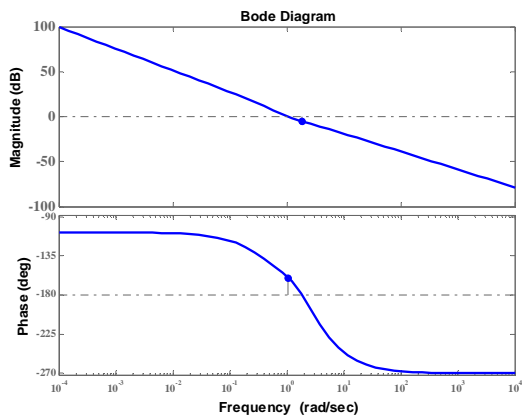

Fig. 5. Open loop Bode plot of controller with the system. Gm = 5db and Pm = 20°

### D. Integer Order (IO) Relay feedback PI controller design

Åström and Hägglund [1] suggest the relay feedback test to generate sustained oscillation as an alternative to the conventional continuous cycling technique. It is very effective in determining the ultimate gain and ultimate frequency. Considering the relay feedback design method in which the switch on point and switch off point are considered as 0.7 and 0.3 respectively in order to generate ultimate period $P_u$. The ultimate gain $K_u$ is calculated with the help of height of the relay (h) and amplitude of oscillation (a).

$$K_u = \frac{4h}{\pi a} \qquad \ldots\ldots\ldots\ldots(21)$$

The values obtained from the responses shown in Figure 6. (a) and (b) are a = 0.2, h = 0.5 and $P_u$ = 2.4. On substituting the values of a and h in equation (21) we got $K_u$ = 3.18. The values of $K_u$ and $P_u$ are used in closed loop Ziegler–Nichols PID controller tuning rule and the PI controller parameters are obtained as $K_c$ = 1.431 and $K_I$ = 0.72. The transfer function of the PI controller is given as

$$G_C(s) = 1.431 + 0.72/s \qquad \ldots\ldots\ldots\ldots(22)$$

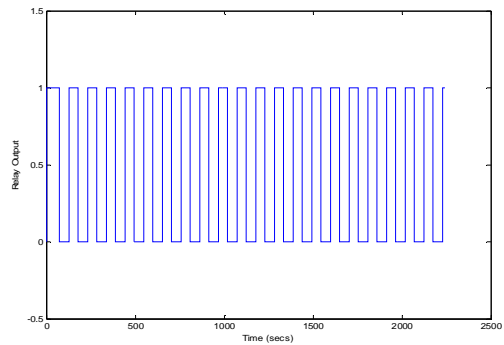

(a)

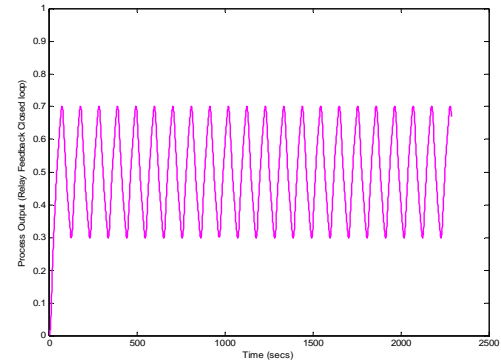

(b)

Fig. 6 (a) Relay output (b) Closed – loop feedback output

Thus the Integer Order Relay Feedback PI (IO-RFPI) controller is designed and controller settings are obtained. In the next section, the fractional order $PI^\lambda$ controller and integer order relay feedback PI controller is compared.

## V. RESULTS AND DISCUSSIONS

### A. Set point Tracking

The performance of set point tracking for the Fractional Order $PI^\lambda$ controller (FO- $PI^\lambda$) and Integer Order Relay Feedback PI controller (IO-RFPI) are simulated as shown in Figure 7 and are made from 50% set point of speed. For each control systems, the performance analysis in the sense of ISE and IAE is compared and tabulated in Table II and Table III.

### B. Load Tracking

The performance of load tracking for the Fractional order $PI^\lambda$ controller (FO-$PI^\lambda$) and Integer Order Relay feedback PI controller (IO-RFPI) are simulated as shown in Figure 9 and are made from 50% set point of speed. For each control systems, the performance analysis in the sense of ISE and IAE is compared and tabulated in Table IV. From the servo and regulatory responses, it is observed that the settling time and rise time of the Fractional order $PI^\lambda$ controller is very much lesser than the Integer order Relay Feedback PI controller. The controller output responses (Armature Voltage) for both the servo and regulatory are shown in Figure 8 and 10 respectively.

TABLE II. ISE AND IAE PERFORMANCE ANALYSIS OF THE FO- $PI^\lambda$ CONTROL SYSTEMS (SERVO).

| FO- $PI^\lambda$ | +5% | +10% | +15 | -5% | -10% | -15% |
|---|---|---|---|---|---|---|
| ISE | 8.77 | 34.84 | 78.34 | 8.90 | 35.10 | 78.73 |
| IAE | 5.34 | 10.60 | 14.8 | 5.36 | 9.65 | 13.3 |

TABLE III. ISE AND IAE PERFORMANCE ANALYSIS OF THE IO- RFPI CONTROL SYSTEMS (SERVO).

| IO-RFPI | +5% | +10% | +15 | -5% | -10% | -15% |
|---|---|---|---|---|---|---|
| ISE | 20.1 | 80.35 | 180.4 | 20.07 | 80.28 | 180.6 |
| IAE | 9.65 | 19.81 | 28.84 | 9.74 | 19.14 | 29.02 |

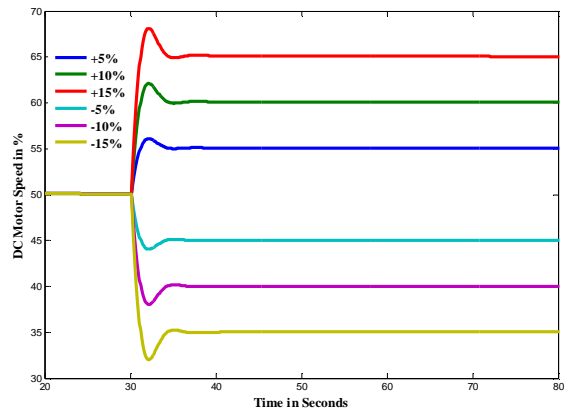

(a)

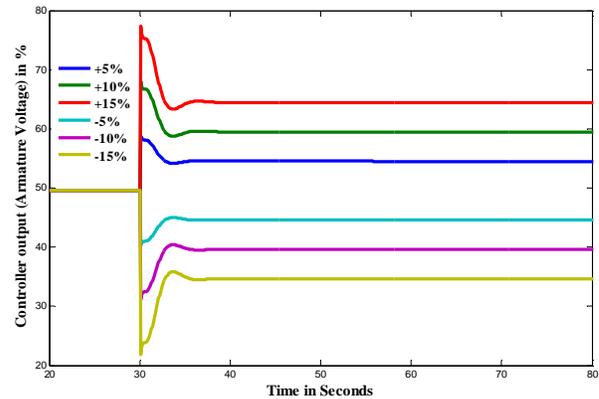

(a)

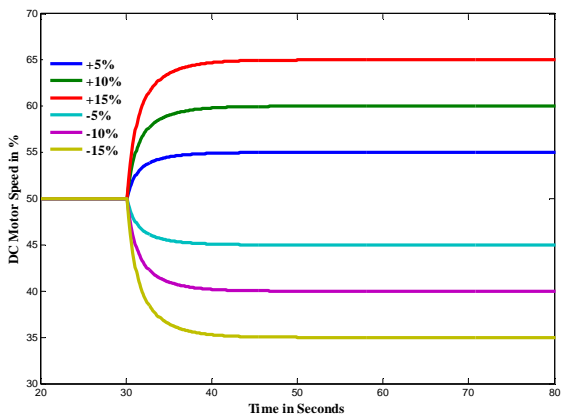

(b)

Fig. 7 Set point tracking performances of FO- $PI^\lambda$ and IO-RFPI

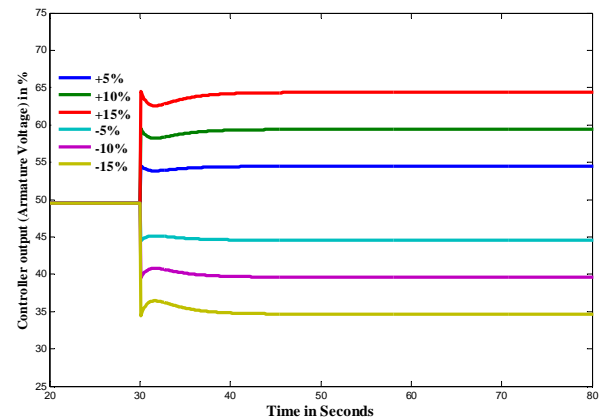

(b)

Fig . 8 Set point tracking - Controller output performances of (a) FO- $PI^\lambda$ and (b) IO-RFPI

Thus the servo response of the Fractional order $PI^\lambda$ controller (FO- $PI^\lambda$) and Integer Order Relay Feedback PI (IO-RFPI) controller are simulated from the 50% rated speed (i.e. 750 rpm) of DC motor. The process output (speed in %) and the controller output (Armature Voltage in %) are recorded.

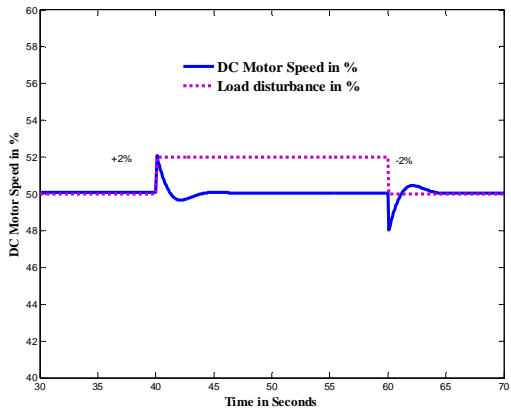

(a)

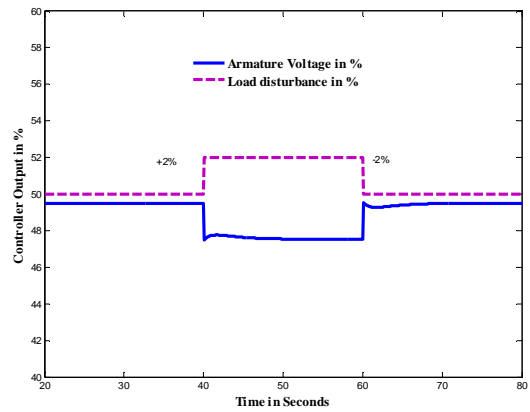

(b)

Fig. 10 Load tracking – Controller performances of
(a) FO- PI² and (b) IO-RFPI

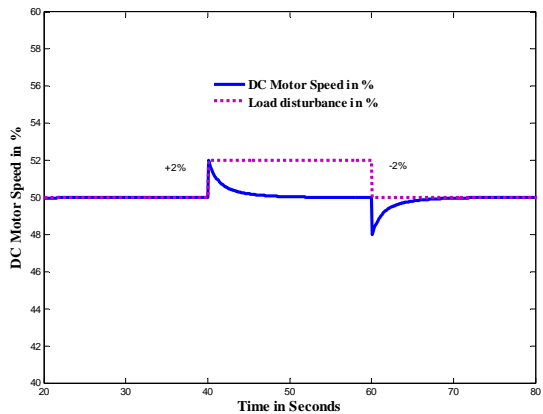

(b)

Fig. 9 Load tracking performances of (a) FO- PI² and (b) IO-RFPI

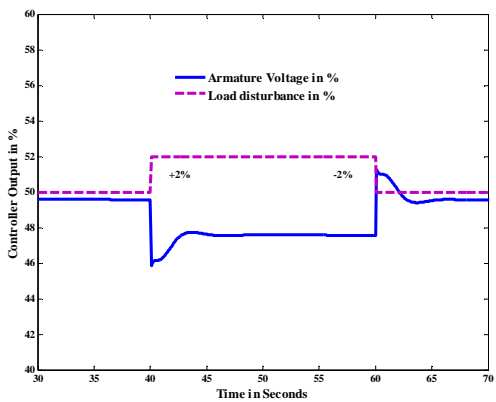

(a)

TABLE IV. ISE AND IAE PERFORMANCE ANALYSIS OF THE FO- PI² AND IO
– RFPI CONTROL SYSTEMS (REGULATORY).

| Control System | ISE | | | | IAE | | | |
|---|---|---|---|---|---|---|---|---|
| **FO- PI²** | +2% | 1.6 | -2% | 1.6 | +2% | 1.9 | -2% | 2.1 |
| **IO - RFPI** | +2% | 3.4 | -2% | 3.4 | +2% | 3.8 | -2% | 3.8 |

Thus the regulatory response of the Fractional order PI² controller (FO- PI²) and Integer Order Relay Feedback PI (IO-RFPI) controller are simulated from the 50% rated speed (i.e. 750 rpm) of DC motor. The process output (speed in %) and the controller output (Armature Voltage in %) are recorded.

*D. Stability Analysis*

The stability of a Fractional Order Control System (FOCS) is analyzed with Matignon's stability theorem [11]. It is well known from the general stability theory that a linear time-invariant (LTI) system is stable if all roots of characteristic equation are negative or have negative real parts. It means that they are located on the left half of the complex plane.

In the fractional-order LTI case, the left half of the plane is different from the integer one. As can be shown in Fig. 11, the vertical axis of the complex plane is changed with an angle depending on the fractional order. Therefore, the stability region for the closed loop poles can be increased or decreased. It should be noted that only the denominator is meaningful for the stability assessment.

The pole position plot of the FOCS obtained using the MATLAB is shown in Fig. 12. This figure shows that all poles of the FOCS are in the fractional left half plane and thus the FOCS is stable. Similarly, the pole position plot for the Integer Order Control System (IOCS) in root locus is shown in Fig. 13 and is stable.

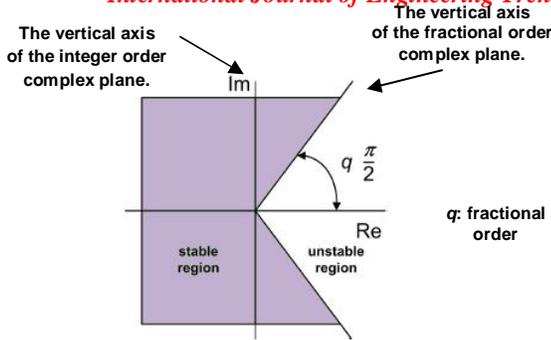

Fig. 11 Stability region of a LTI FO system with order 0 <q ≤ 1

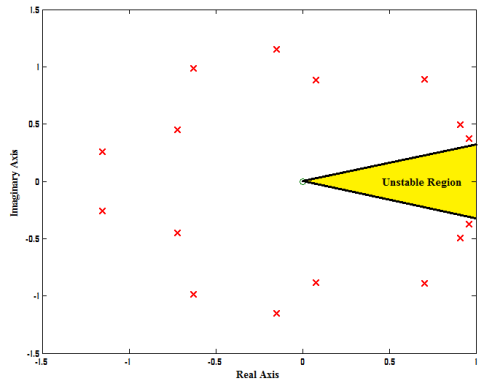

Fig. 12 The poles of the fractional order control system.

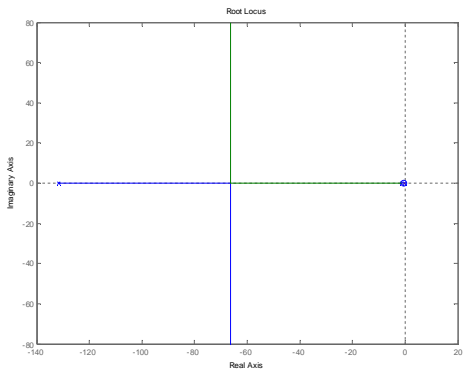

Fig. 13 The poles of the integer order control system.

Thus the stability of FO- PI$^\lambda$ and (IO-RFPI) controller are simulated and recorded. It is found that both the systems are stable.

## VI. CONCLUSIONS

In this paper, a fractional order PI$^\lambda$ controller is designed based on global stability boundary locus method. The controller parameters are obtained based on providing the desired gain margin (GM) and phase margin (PM) in the global stability region. It is seen from the simulation and performance analysis results that the fractional order PI$^\lambda$ controller shows better results compared to that of relay feedback PI controller for the DC motor speed control system. The stability of the both control systems is analyzed and is stable.